\documentclass[12pt]{article}

\usepackage[T1]{fontenc}
\usepackage{graphicx}
\usepackage{mathrsfs}
\usepackage{amsmath}
\usepackage{amssymb}
\usepackage{stackrel}
\usepackage{MnSymbol}

\newfont{\mbld}{cmbx10 scaled 800}
\newfont{\cab}{cmsy10 scaled 1200}
\newfont{\scab}{cmsy10 scaled 1000}
\newfont{\bcall}{cmbsy10 scaled 1200}

\def\XXint#1#2#3{{\setbox0=\hbox{$#1{#2#3}{\int}$ }
\vcenter{\hbox{$#2#3$ }}\kern-.65\wd0}}

\begin{document}
\title{\Large On a singular Fredholm--type integral equation
\\
\Large arising in ${\cal N}=2$ super--Yang--Mills theories}
\author{\normalsize Franco Ferrari$^1$\footnote{e-mail:
    ferrari@fermi.fiz.univ.szczecin.pl}
and Marcin Pi\c{a}tek$^{1,2}$\footnote{e-mail:
  piatek@fermi.fiz.univ.szczecin.pl} \\
\normalsize $^1$ Institute of Physics and CASA*, University of Szczecin,\\
\normalsize Wielkopolska 15, 70451 Szczecin, Poland\\
\normalsize $^2$ Bogoliubov Laboratory of Theoretical Physics,\\
\normalsize Joint Institute for Nuclear Research, 141980, Dubna, Russia
}
\maketitle
\abstract{
In this work we study the Nekrasov--Shatashvili limit of the Nekrasov
instanton partition function of  Yang--Mills
field theories with ${\cal N}=2$ supersymmetry and gauge group
$SU(N_c)$. The theories are coupled with $N_f$ flavors of fundamental
matter. The 
equation that determines the density of eigenvalues at the leading
order in the saddle-point approximation is
exactly solved when $N_f=2N_c$. The dominating contribution to the
instanton free energy is 
computed. The requirement that this energy is finite imposes
quantization conditions on the parameters of the theory that
are in agreement with analogous  conditions that have been
derived in previous works.
The instanton energy and
thus the instanton contribution to the prepotential of the gauge
theory is computed in closed form.
}

\section{Introduction}
In 1994 Seiberg and Witten proposed an exact solution for  low-energy
effective ${\cal N}=2$ supersymmetric gauge theories \cite{SW1,SW2}.
This solution has led to considerable  progress  in
understanding the strong coupling dynamics of gauge theories.
In recent days supersymmetric Yang--Mills theories with ${\cal N}=2$
supersymmetry 
have become again a subject of intense studies
mainly due to the  discovery of the so-called AGT--W \cite{AGT,Wyllard}
and Bethe/gauge correspondences \cite{NekraSha,NekraSha2,NekraSha3}.
A crucial ingredient of these conjectures is the multi-instanton
Nekrasov partition function $Z_{\sf inst}(q,\epsilon_1,\epsilon_2)$
\cite{Nekrasov1,NekrasovOkounkov,NekraShadchin,Marshakov1} that takes into account
the instanton sector
of classes of four dimensional quiver gauge field theories with ${\cal N}=2$ supersymmetry.
The ${\cal N}=2$ gauge theory must be embedded in a
$\Omega-$background in order to apply the
localization technique which allows
to express $Z_{\sf inst}(q,\epsilon_1,\epsilon_2)$ in the form of a sum
of contour integrals over $k$ variables $\phi_I$, $I=1,\ldots,k$,
where the integer $k$ with $0\le k<+\infty$ denotes the instanton number.
Moreover, $q$ is an effective scale, while $\epsilon_1$ and
$\epsilon_2$ are the 
deformation parameters of the $\Omega-$background.

For a wide set of applications 
\cite{NekraSha,Pogho1,Fucitoiinni,BMT,Dorey,Dorey2,Piatek1,FFPiatek},
it is sufficient to evaluate
$Z_{\sf inst}(q,\epsilon_1,\epsilon_2)$
  in the Nekrasov-Shatashvili limit
$\epsilon_2\to 0$ while $\epsilon_1$ is kept finite.
This limit is relevant for the computation of the so-called effective
twisted superpotential \cite{NekraSha}, which determines the
low energy effective dynamics of two-dimensional gauge theories
restricted to the $\Omega-$background. The twisted superpotential
plays also a pivotal role in the already mentioned Bethe/gauge
correspondence that maps supersymmetric vacua of the ${\cal N}=2$
theories to Bethe states of quantum integrable systems. The result of
that duality is that the twisted superpotential is identified with the
Yang's functional \cite{NekraSha} which describes the spectrum of the
corresponding quantum integrable system. In the AGT correspondence,
the limit $\epsilon_2\to 0$ while $\epsilon_1$ remains finite allows
to study the semi-classical limit of the Liouville field theory
\cite{Piatek1,FFPiatek}. 

The Nekrasov-Shatashvili limit can be investigated with the
saddle-point method, which 
requires the 
solution of an infinite system of algebraic equations in the discrete variables
$\phi_I$ mentioned above (cf. \cite{Pogho1,Fucitoiinni}). This task
is usually accomplished after applying some approximation, for instance by
performing a series expansion in the parameter $q$ and computing the
coefficients of the expansion by a recursive algorithm
\cite{Pogho1,Fucitoiinni}. Alternatively, by the
introduction of the density
function that describes the distribution of the $\phi_I$'s,
the saddle point equations may be cast in the form of a Fredholm
equation on the continuous line. The
density function becomes in fact  continuous in the limit
$\epsilon_2\to
0$. Up to now, these Fredholm equations have
been only marginally investigated, see for example
\cite{Pogho1,Fucitoiinni}. The problem of finding their
exact solutions has not been studied so far. This will be the
subject of the
present work. To fix the ideas, we consider here the case of a
${\cal N}=2$ supersymmetric Yang--Mills theory with $SU(N_c)$ gauge
group of local symmetry. The theory is coupled to matter in the
fundamental representation and with a number $N_f$ of flavors.
The Fredholm equation that determines the configurations of the
density functional at the leading order in the saddle-point approximation
is solved in closed form when $N_f=2N_c$ without the need of
performing a series 
expansion in the parameter $q$. 
The particular case  $N_f=2N_c$ 
includes
the
entire class of conformal ${\cal N}=2$ gauge field theories with
vanishing $\beta-$function. It
is also relevant in order to
compute the classical limit of the quantum 4-point Virasoro conformal
block of the Liouville field theory exploiting the AGT correspondence.

Using the exact solution of the saddle-point equation mentioned above,
the
expression of the instanton energy is evaluated. It is found that the
consistency requirements that are necessary in order to keep the
energy finite
impose a quantization condition on the masses of the fundamental matter
and on the vacuum
expectation values of the adjoint scalar fields.
This condition is in agreement with analogous conditions that have
been  postulated in previous works, see for example \cite{Dorey,Dorey2}.

The material of this paper is divided as follows. In the
next Section the instanton partition function
$Z_{\sf inst}(q,\epsilon_1,\epsilon_2)$ is studied in the
thermodynamic limit, i.e.
when $\epsilon_2\to 0$ while $\epsilon_1$ is kept finite.
Following the arguments of \cite{NekrasovOkounkov} and statistical mechanics,
it is shown that the dominant number $\bar k$ of instantons is $\bar k=\frac
X{\epsilon_2}$,
where $X$ is a constant that will be determined later in
Section 3. This constant is physically important because it is directly
related  to the instanton corrections to the prepotential of the gauge
theory and, via the Matone's identity \cite{Matone}, to the instanton
corrections of
the scalar field condensate.  A field theoretical expression of the
instanton partition function in the thermodynamic limit is provided.
In Section 3 the saddle point equation is solved exactly when
$N_f=2N_c$. In order 
to determine the value of the $X$ parameter mentioned above and the
length of the interval in which the one--cut solutions are defined, a
set of two
implicit algebraic equations is
provided. The instanton free energy is evaluated exactly up to the
solution of those equations.
To verify the consistency of our calculations, 
we consider the differential equation that has been used in
\cite{Fucitoiinni} to compute the instanton contribution to the
prepotential of the ${\cal 
  N}=2$ gauge field theory.
We check that the instanton contribution to the prepotential computed
here in closed form satisfies the same differential equation.
Finally, our conclusions are drawn in Section 5.

\section{Continuous density representation}
In this work we consider the instanton partition function of a ${\cal
  N}=2$ gauge field theory with gauge group $SU(N_c)$ and matter in the
fundamental representation \cite{NekrasovOkounkov,Fucitoiinni}:
\begin{eqnarray}
Z_{\sf inst}(q,\epsilon_1,\epsilon_2)&=&
1+\sum_{k=1}^{\infty}\frac{q^k}{k!}\left(\frac{\epsilon_1+\epsilon_2}{
  \epsilon_1\epsilon_2}\right)^k
\int\limits_{\mathbb{R}}\prod_{I=1}^k\frac{d\phi_I}{2\pi
i}\prod_{I\ne J=1}^kD(\phi_I-\phi_J)
\nonumber\\
&\times&
\prod_{I=1}^kQ(\phi_I)
\label{zinst}
\end{eqnarray}
where
$D(z)=\frac{z(z+\epsilon_1+\epsilon_2)}{(z+\epsilon_1)(z+\epsilon_2)}$,
$Q(z)=\frac{M(z)}{P(z+\epsilon_1+\epsilon_2)P(z)}$,
$M(z)=\prod_{r=1}^{N_f}(z+m_r)$
and
$P(z)=\prod_{l=1}^{N_c}(z-a_l)$.
Here the $a_l$'s, $l=1,\ldots,N_c$, are the vacuum expectation values of
the adjoint scalar field in the $SU(N_c)$ vector multiplet, while the
$m_r$'s parametrize the masses of the fundamental matter.
$N_f$ can be indentified with the number of flavors of the theory,
though a more precise description of its meaning can be found in
\cite{Nekrasov1}. Let us note that the integrals over the real line
$\mathbb{R}$ in Eq.~(\ref{zinst}) require some form of regularization
and should be intended as integrals over a closed contour. The details
are explained in \cite{Nekrasov1,Parisi}.

In the following, we will study the partition function
$Z_{\sf inst}(q,\epsilon_1,\epsilon_2)$ in the ``thermodynamic'' limit
$\epsilon_2=0$ that,
according to \cite{NekraSha}, is related to the quantization
of classical systems.

For very small values of $\epsilon_2$, Eq.~(\ref{zinst}) reduces to:
\begin{equation}
Z_{\sf inst}(q,\epsilon_1,\epsilon_2)\sim 1+\sum_{k=1}^{\infty}\frac
1{k!}\int\limits_{\mathbb{R}} \prod_{I=1}^k\frac{d\phi_I}{2\pi i\,\epsilon_2}\,\textrm{e}^{\frac
1{\epsilon_2}W_k(\{\phi_I\})}
\end{equation}
where
\begin{equation}
W_k(\{\phi_I\})=\sum_{I>J=1}^k\epsilon_2^2G(\phi_I-\phi_J)+
\sum_{I=1}^k\epsilon_2\log(q Q_0(\phi_I))\label{wkphiI}
\end{equation}
with
\begin{equation}
G(\phi_I-\phi_J)=\frac{2\epsilon_1}{\epsilon_1^2-(\phi_I-\phi_J)^2}
\label{gphidef}
\end{equation}
and
$Q_0(z)=\frac{M(z)}{P(z+\epsilon_1)P(z)}$.
 According to \cite{NekrasovOkounkov}, in the thermodynamic limit
$\epsilon_2\to0$
the instantonic partition function is dominated by the contributions
for which
\begin{equation}
k=\bar k= \frac X{\epsilon_2}\label{domvaluek}
\end{equation}
where $X$ is a finite proportionality constant. As a consequence,
$Z_{\sf inst}(q,\epsilon_1,\epsilon_2)$ may be approximated as follows:
\begin{equation}
Z_{\sf inst}(q,\epsilon_1,\epsilon_2)\underset{\epsilon_2\to 0}{\sim}
Z_{\bar k}(q,\epsilon_1,\epsilon_2)=\int\limits_{\mathbb{R}}\prod_{I=1}^{\bar k}d\phi_I
\,\textrm{e}^{\frac 1{\epsilon_2}\,W_{\bar k}(\{\phi_I\})}\label{thermlim}
\end{equation}
with $W_{\bar k}(\{\phi_I\})$ being defined in Eq.~(\ref{wkphiI}).
From now on, we will consider the instantonic partition function in
the thermodynamic limit (\ref{thermlim}).

Following \cite{ItzyksonDrouffe,MMarino}, we perform the continuous
limit of the instanton partition function by introducing the new
field variable $\phi(I\epsilon_2)$ defined by the relations:
$\phi_I=\phi(I\epsilon_2)$, $I=1,\ldots,\bar k$.
Of course, when $I=\bar k$ we have that $I\epsilon_2=X$. We note also that
the quantity $x_I=I\epsilon_2$ may be considered as a discrete
coordinate in the interval
$[0,X]$. It is easy to realize that in terms of the field $\phi(x_I)$
the instantonic partition function becomes:
\begin{equation}
Z_{\sf inst}(q,\epsilon_1,\epsilon_2)
\underset{\epsilon_2\to 0}{\sim}
\int\limits_{\mathbb{R}}\prod_{I=1}^{\bar
  k}\frac{d\phi(x_I)}{2\pi i}\exp\left[
\frac 1{\epsilon_2}\,W_{\bar k}(\{\phi(x_I)\})
\right]\label{zkxI}
\end{equation}
with
\begin{equation}
W_{\bar k}(\{\phi(x_I)\})=\sum_{I<J=1}^{\bar
  k}\epsilon_2^2G(\phi(x_I)-\phi(x_J))+
\sum_{I=1}^{\bar k}\epsilon_2\log(qQ_0(\phi(x_I)))
\end{equation}
If $\epsilon_2$ is vanishingly small, $x_I$ may be replaced by the
continuous variable $x\in[0,X]$ and the expression of
$Z_{\sf inst}(q,\epsilon_1,\epsilon_2)$ may
formally be written as follows:
\begin{equation}
Z_{\sf inst}(q,\epsilon_1,\epsilon_2\sim 0)=\int{\cal
  D}\phi(x)\exp\left[ 
\frac 1{\epsilon_2}W_{\bar k}(\phi(x))
\right]\label{zkx}
\end{equation}
where
$
{\cal D}\phi(x)=\prod_{I=1}^{\bar k}\frac{d\phi(x_I)}{2\pi i\,\epsilon_2}
$
and
\begin{equation}
W_{\bar k}(\phi(x))=\int_0^Xdx\int_0^xdyG(\phi(x)-\phi(y))
+\int_0^Xdx\log(qQ_0(\phi(x)))
\label{wkphix}
\end{equation}
In the integrals over $x$ and $y$
of Eq.~(\ref{wkphix})
the principal value prescription is
implicitly understood in order to cure the singularity of $G(z)$ at
the points $z=\pm \epsilon_1$.
\section{Saddle point evaluation}
For small values of $\epsilon_2$, the summation over all
configurations $\phi(x)$ in Eq.~(\ref{zkx}) will be dominated
by those configurations $\phi_{\sf cl}(x)$ that correspond to the absolute
maximum of the $\bar k-$instantonic energy $W_{\bar k}(\phi(x))$, i.~e.:
\begin{equation}
Z_{\sf inst}(q,\epsilon_1,\epsilon_2)\sim \exp\left[
\frac 1{\epsilon_2}\, W_{\bar k}(\phi_{\sf cl}(x))
\right]\label{phicldef}
\end{equation}
In order to find the extremal points of the instantonic energy, we
introduce the density $\rho(\phi)$ defined by the relation:
\begin{equation}
d\phi\rho(\phi)=dx
\label{rhophirel}
\end{equation}
Eq.~(\ref{rhophirel}) implies also the condition:
\begin{equation}
\int_{-\infty}^{+\infty}d\phi\rho(\phi)=X
\label{normalcond}
\end{equation}
With the help of (\ref{rhophirel}), the functional $W_{\bar
  k}(\phi(x))$ of Eq.~(\ref{wkphix})
can be expressed as a functional of $\rho(\phi)$:
\begin{eqnarray}
\textrm{e}^{\frac 1{\epsilon_2}\,W_{\bar k}(\phi(x))}&=&
\exp\left[\frac 1{\epsilon_2}\left(
\frac12\,
\strokedint_{-\infty}^{+\infty}d\phi\,\strokedint_{-\infty}^{+\infty}
d\phi'\rho(\phi)\rho(\phi')G(\phi-\phi')\right.\right.\nonumber\\
&+&\left.\left.\strokedint_{-\infty}^{+\infty}d\phi\rho(\phi)\log(qQ_0 (\phi))
\right)
\right]\label{passagetorho}
\end{eqnarray}
where
$\strokedint_{-\infty}^{+\infty}f(\phi)d\phi$
denotes the principal value of the integral of a generic function
$f(\phi)$ with
singularities to be cured with the principal
value prescription located
at finite points and/or at infinity.
In order to obtain Eq.~(\ref{passagetorho}) from
Eq.~(\ref{wkphix}) we have used the identity:
\begin{equation}
\int_0^Xdx\int\limits_0^xdyG(\phi(x)-\phi(y))=\frac
12\int_0^Xdx\int_0^XdyG(\phi(x)-\phi(y))
\label{ideone}
\end{equation}
The stationary configurations of the exponent present in the right
hand side of 
Eq.~(\ref{passagetorho}) are given by the solutions of the saddle point
equation:
\begin{equation}
\strokedint_{-\infty}^{+\infty}d\phi'\rho(\phi')G(\phi-\phi')+\log(qQ_0(\phi))=0
\label{eqmots}
\end{equation}
Once a solution $\rho_{\sf cl}(\phi)$ of the above equation is found,
the corresponding configurations $\phi_{\sf cl}(x)$ appearing in
Eq.~(\ref{phicldef})
that maximize the energy $W_{\bar k}(\phi_{\sf cl}(x))$ may be recovered
using Eq.~(\ref{rhophirel}).
To solve Eq.~(\ref{eqmots}) it is possible to exploit the
superposition principle and split $\rho(\phi)$ as follows:
\begin{equation}
\rho(\phi)=\rho_q(\phi)+\rho_{Q_0}(\phi)\label{split}
\end{equation}
where $\rho_q(\phi)$ and $\rho_{Q_0}(\phi)$ are respectively the
solutions of the equations:
\begin{equation}
\strokedint\limits_{-\infty}^{+\infty}
d\phi'\rho_q(\phi')G(\phi-\phi')+\log(q)=0\label{rhoq}
\end{equation}
and
\begin{equation}
\strokedint\limits_{-\infty}^{+\infty}
d\phi'\rho_{Q_0}(\phi')G(\phi-\phi')+\log(Q_0(\phi))=0
\label{rhoQzero}
\end{equation}
Let us consider Eq.~(\ref{rhoq}) first.
Following \cite{FWKing}, a solution of this equation can exist only if
the support of 
$\rho_q(\phi)$ is limited to some finite interval $[-L,L]$ on the
$\phi-$axis. The value of $L$
will be fixed later.
After some calculations, which will not be reported here because
are similar to those reported in
\cite{FWKing}, Sections 11.4 and 12.2, one
obtains:
\begin{equation}
\rho_q(\phi)=-\frac{\log (q)
}{2\epsilon_1\pi}\sqrt{L^2-\phi^2}[\theta(L-\phi) -\theta(-L-\phi)]
\label{calcrhoq}
\end{equation}
where $\theta(x)$ denotes the Heaviside theta function.

At this point we are ready to solve Eq.~(\ref{rhoQzero}). 
To this
purpose, it is convenient to pass to the Fourier representation.
The problem
is to evaluate
the Fourier
transform of  $\log(Q_0(\phi))$, since this function contains
logarithms
and the Fourier transform of the logarithm
\cite{distributionbooks,distributionbooks2} is not
uniquely defined
\cite{refonlogs}. To avoid
this problem, we have to
assume the condition:
\begin{equation}
N_f=2N_c
\label{assumptionone}
\end{equation}
Under this requirement, the function $\log(Q_0(\phi))$ has no longer
singularities at infinity and its Fourier transform can be determined
without ambiguities.
The condition (\ref{assumptionone}) covers the case of ${\cal N}=2$ gauge
field theories with $SU(N_c)$ group of symmetry which are conformal --
their $\beta-$function vanishes identically. Our main motivation to
study theories with $N_f=2N_c$ comes from the AGT correspondence that
relates the 
4-point conformal block of the Liouville field theory with the
$SU(2)$, $N_f=4$
Nekrasov instanton partition function $Z_{\sf
  inst}^{SU(2),N_f=4}$. Exploiting such correspondence, one finds that
the $\epsilon_2\to0$ limit of  $Z_{\sf
  inst}^{SU(2),N_f=4}$ is related to the classical limit of the
quantum 4-point conformal block \cite{Piatek1,FFPiatek}.

Denoting with $\tilde{\rho}_{Q_0}(\omega),\tilde G(\omega)$ and
$\lambda(\omega)$ respectively the Fourier transforms of
$\rho_{Q_0}(\phi)$, $G(\phi-\phi')$  and $\log(Q_0(\phi))$, we obtain
from Eq.~(\ref{rhoQzero}) the following identity:
\begin{equation}
\sqrt{2\pi}\tilde\rho_{Q_0}(\omega)\tilde
G(\omega)=-\lambda(\omega)
\label{rhoQzeroFT}
\end{equation}
with:
\begin{equation}
\tilde
G(\omega)=+\sqrt{2\pi}\,\mbox{sgn}\omega\sin(\epsilon_1\omega)\label{Gtildek}
\end{equation}
and
\begin{equation}
\lambda(\omega)=\left(\frac\pi 2\right)^{\frac 12}\frac
1\omega\,\mbox{sgn}\omega \left(
\sum_{l=1}^{N_c} \textrm{e}^{i\omega a_l}+\sum_{l=1}^{N_c}
\textrm{e}^{i\omega (a_l-\epsilon_1)} 
-\sum_{r=1}^{N_f} \textrm{e}^{-i\omega m_r}
\right)\label{lambdakdef}
\end{equation}
The function sgn$\omega$ appearing in the two equations above
denotes the sign function and it is defined in such
a way that it is zero when $\omega=0$.
The Fourier transforms of $G(\phi)$ and $\log(Q_0(\phi))$ have been
computed using the following formulas:
\begin{equation}
\int_{-\infty}^{+\infty}
\frac{d\phi}{\sqrt{2\pi}}\, \textrm{e}^{i\omega\phi}
\frac{1}{\phi-a}=\, \textrm{e}^{ia\omega}\left(\frac\pi2 \right)^{\frac
  12}i\mbox{sgn}\omega
\end{equation}
and
\begin{equation}
\int\limits_{-\infty}^{+\infty}\frac{d\phi}{\sqrt{2\pi}}
\,\textrm{e}^{i\omega\phi}\log\left(
\frac{\phi-A}{\phi-B}
\right) =\left(\frac\pi
2\right)^{\frac 12}\frac{\mbox{sgn}\omega}{\omega}
\left( \textrm{e}^{i\omega B}-\textrm{e}^{i\omega
A}\right)
\end{equation}
where $A,B$ are constants.
After substituting Eqs.~(\ref{Gtildek}) and (\ref{lambdakdef}) in
Eq.~(\ref{rhoQzeroFT}), it is easy to derive
$\tilde\rho_{Q_0}(\omega)$
for $\omega\ne 0$:
\begin{equation}
\tilde\rho_{Q_0}(\omega)=\frac
1{2\omega\sin\omega\epsilon_1}\frac
1{\sqrt{2\pi}}\left(\sum_{r=1}^{N_f}\textrm{e}^{-i\omega m_r}-
\sum_{l=1}^{N_c} \textrm{e}^{i\omega a_l}-\sum_{l=1}^{N_c} \textrm{e}^{i\omega (a_l-\epsilon_1)}
\right)\label{rhokdifzero}
\end{equation}
When $\omega=0$, Eq.~(\ref{rhoQzeroFT})
becomes undefined, because $\tilde
G(\omega)$ and $\lambda(\omega)$ are both
proportional to sgn$\omega$ which is
zero when $\omega=0$.
From the normalization condition (\ref{normalcond}), we know that
$\tilde\rho_{Q_0}(0)$ must be finite. As a matter of fact,
Eq.~(\ref{normalcond}) implies:
\begin{equation}
-\frac{L^2\log(q)}{4\epsilon_1}+\sqrt{2\pi}\tilde\rho_{Q_0}(0)=X
\label{normcondtwo}
\end{equation}
In order to  derive the above equation, we have used the fact that
\begin{equation}
\strokedint_{-\infty}^{+\infty}\rho_{Q_0}(\phi)d\phi=\sqrt{2\pi}
\tilde\rho_{Q_0}(0) \label{ftrhoQzero}
\end{equation}
and the relation
$
\strokedint_{-\infty}^{+\infty}\rho_q(\phi)d\phi=-
\frac{L^2\log(q)}{4\epsilon_1}
$.
Due to Eq.~(\ref{normcondtwo}), $\tilde\rho_{Q_0}(0)$ is finite
because $X$ is finite by definition.
On the other side, 
if we consider the expression
of $\tilde\rho_{Q_0}(\omega)$ just computed in
Eq.~(\ref{rhokdifzero}), it is easy to check that  $\lim_{\omega\to
  0}|\tilde\rho_{Q_0}(\omega)|=+\infty$ if the values
of the
parameters $a_l$
and $m_r$ are arbitrary. As a matter of fact, the function
$\tilde\rho_{Q_0}(\omega)$ of Eq.~(\ref{rhokdifzero}) exhibits a
double pole in $\omega=0$ and it is divergent at that point.
Besides, the pole in $\omega=0$ makes $\tilde\rho_{Q_0}(\omega)$ not
integrable with respect to $\omega$ even using the principal value
prescription. 
To eliminate this paradox, in such a way that
Eq.~(\ref{rhokdifzero}) will be consistent with
Eq.~(\ref{normcondtwo}) and
$\tilde \rho_{Q_0}(\omega)$ will be a continuous function in
$\omega=0$, it is necessary the introduction of the following
additional condition on the parameters
$a_l$'s and $m_r$'s:
\begin{equation}
\sum_{r=1}^{N_f}m_r+2\sum_{l=1}^{N_c}a_l=N_c\epsilon_1 .
\label{condone}
\end{equation}
Under this condition, we find from  Eqs.~(\ref{rhokdifzero}) after a
simple expansion near the point $\omega=0$ that:
\begin{equation}
\tilde\rho_{Q_0}(0)=\frac 1{4\epsilon_1\sqrt{2\pi}}\left[
\sum_{l=1}^{N_c}(a_l^2+(a_l-\epsilon_1)^2)-\sum_{r=1}^{N_f}m_r^2
\right]\label{rhoQzeroinzero}
\end{equation}
It is now possible to derive a new relation which, together
with Eq.~(\ref{normcondtwo}), will be able to determine the still
unknown parameters $X$ and $L$.
Indeed, let us multiply both terms of 
Eq.~(\ref{rhoQzero}) by $\rho_q(\phi)$ and integrate over
$\phi$. Applying Eq.~(\ref{rhoq}) in order to eliminate
$\rho_q(\phi)$ from the first term, one may prove the following identity:
\begin{eqnarray}
\tilde\rho_{Q_0}(0)&=&-\frac{1}{(2\pi)^{\frac
    32}\epsilon_1}\int_{-L}^{+L}d\phi' 
\sqrt{L^2-\phi^{\prime2}}\log(Q_0(\phi'))\label{relone}
\end{eqnarray}
Thus, remembering Eq.~(\ref{rhoQzeroinzero}) it is possible to check
that:
\begin{equation}
\int\limits_{-L}^{+L}d\phi' 
\sqrt{L^2-\phi^{\prime2}}\log(Q_0(\phi'))=
-\frac\pi2\left(
\sum_{l=1}^{N_c}(a_l^2+(a_l-\epsilon_1)^2)-\sum_{r=1}^{N_f}m_r^2
\right)
\label{reltwo}
\end{equation}
Eq.~(\ref{reltwo}) provides an implicit equation which allows to
derive the value of $L$ as a function of the parameters of the ${\cal
  N}=2$ gauge field theory.
Once the value of $L$ is known,  that
of $X$ can be determined thanks to the normalization condition
(\ref{normcondtwo}). An important consequence of Eq.~(\ref{reltwo}) is
that $L$ does not depend on $q$. This fact will become relevant at
the
end of this Section, when we will have to compute the derivative of
the instanton energy with respect to that variable.

As a final remark concerning the expression of  $\tilde
\rho_{Q_0}(\omega)$, we note in Eq.~(\ref{rhokdifzero})
the presence of additional singularities at the points
\begin{equation}
\omega=\frac{\pi\sigma}{\epsilon_1}\qquad\qquad
\sigma=\pm1,\pm2,\ldots\label{polessin}
\end{equation}
They appear due to the presence of the function
$\sin(\omega\epsilon_1)$ in the denominator of
Eq.~(\ref{rhokdifzero}).
These singularities do not pose problems and, whenever it is necessary to
perform an integration involving $\tilde
\rho_{Q_0}(\omega)$, they can be cured with the
help of the principal value prescription. There is an infinite number
of these singularities, but their contributions after taking the
principal value prescription amount to a series which is convergent.
The convergence of  $\tilde
\rho_{Q_0}(\omega)$ allows the computation of $\rho_{Q_0}(\phi)$ by
applying the inverse Fourier transform: $\rho_{Q_0}(\phi)=\frac
1{\sqrt{2\pi}} \int_{-\infty}^{+\infty}d\omega e^{-i\omega\phi}\tilde
\rho_{Q_0}(\omega)$.

Next, we are going to compute the energy
$W_{\bar k}(\phi(x))$ appearing in the right hand side of
Eq.~(\ref{passagetorho}). 
Since we restrict ourselves to the stationary configurations
$\rho(\phi)$ satisfying the saddle point equation (\ref{eqmots}), it
is possible to rewrite $W_{\bar k}(\phi(x))$ in the more compact form:
\begin{equation}
W=\frac 12
\,\strokedint\limits_{-\infty}^{+\infty}d\phi\rho(\phi)\left(
\log q+\log(Q_0(\phi))
\right)\label{Walttwo}
\end{equation}
where we have put $W_{\bar k}(\phi(x))=W$ for simplicity.
Taking into account also the splitting of the density $\rho(\phi)$
made in
Eq.~(\ref{split}) and the fact that at the stationary point the
following identity holds:
$
\strokedint_{-\infty}^{+\infty}d\phi\rho_q(\phi)\log(Q_0(\phi))=\log q
\,\strokedint_{-\infty}^{+\infty}d\phi\rho_{Q_0}(\phi)
$,
we arrive at an expression of $W$ which will be convenient for further
calculations:
\begin{equation}
W=W_{qq}+W_{qQ_0}+W_{Q_0Q_0}\label{instantonenergy}
\end{equation}
In the above equation we have put
$
W_{qq}=\frac {\log
  q}2\,\strokedint\limits_{-\infty}^{+\infty}d\phi\rho_q(\phi)$ and
\begin{equation}
W_{qQ_0}=\log
q\,\strokedint\limits_{-\infty}^{+\infty}d\phi\rho_{Q_0}(\phi)\qquad
W_{Q_0Q_0}=\frac
12\,\strokedint\limits_{-\infty}^{+\infty}d\phi\rho_{Q_0}(\phi)\log(Q_0(\phi))
\label{WQzeroQzero}
\end{equation}
A straightforward computation shows that:
\begin{equation}
W_{qq}=-\frac{L^2(\log q)^2}{8\epsilon_1}
\label{wqqfinal}
\end{equation}
$W_{qQ_0}$ may be derived remembering that
$
\strokedint\limits_{-\infty}^{+\infty}d\phi\rho_{Q_0}
(\phi)=\sqrt{2\pi}\tilde\rho_{Q_0}(0)
$
due to Eq.~(\ref{ftrhoQzero}).
The expression of $\tilde\rho_{Q_0}(0)$ is already known from
Eq.~(\ref{rhoQzeroinzero}), so that it is possible to write:
\begin{equation}
W_{qQ_0}=\frac{\log q}{4\epsilon_1}\left[
\sum_{l=1}^{N_c}\left(a_l^2+(a_l-\epsilon_1)^2
\right)-\sum_{r=1}^{N_f}m_r^2
\right]
\label{wqQzerofinal}
\end{equation}
$W_{qq}$ and $W_{qQ_0}$ are the most important contributions to the
energy from the physical point of view because they contain the
parameter $q$.
Finally,  $W_{Q_0Q_0}$ is given by:
\begin{eqnarray}
W_{Q_0Q_0}&=&
\strokedint\limits_{-\infty}^{+\infty}\frac{d\omega}{\sqrt{2\pi}}
\frac{\lambda(-\omega)}{
4\omega\sin\omega\epsilon_1}\nonumber\\
&\times& \left(\sum_{r=1}^{N_f}\textrm{e}^{-i\omega m_r}-
\sum_{l=1}^{N_c} \textrm{e}^{i\omega a_l}-\sum_{l=1}^{N_c} \textrm{e}^{i\omega (a_l-\epsilon_1)}
\right)\label{wQzeroQzerofinal}
\end{eqnarray}
where $\lambda(-\omega)$ can be obtained from Eq.~(\ref{lambdakdef})
after inverting the sign of $\omega$ in the right hand side.
The potential singularities of the integrand
at $\omega=0$ have been removed by the condition (\ref{condone}),
while the principal value prescription is used to cure the poles at
the points defined by Eq.~(\ref{polessin}).

We are now able to compute the instanton contribution to the
prepotential ${\cal F}$ of the gauge field theory. 
In our settings, $\cal F$ is given
for small values of $\epsilon_1$ by:
\begin{equation}
{\cal F}=-\epsilon_1 W
\end{equation}
with $W$ being the instanton energy
computed 
in Eq.~(\ref{instantonenergy})
using the stationary configurations of $\rho(\phi)$ at the
thermodynamic limit.
From Eq.~(\ref{instantonenergy}) it turns out that
\begin{equation}
{\cal F}=-\epsilon_1 (W_{qq}+W_{q Q_o}+W_{Q_0 Q_0})\label{prepo}
\end{equation}
The expressions of
$W_{qq},W_{q Q_o}$ and $W_{Q_0 Q_0}$
are given in Eqs.~(\ref{wqqfinal}),
(\ref{wqQzerofinal}) and (\ref{wQzeroQzerofinal}) respectively.
The only quantity that is not explicitly known is the
parameter $L$ appearing in Eqs.~(\ref{wqqfinal}).
This may be  obtained from Eq.~(\ref{reltwo}) exploiting numerical
techniques. 

To check the consistency of our calculations, we verify that the
expression of $\cal F$ derived here satisfies the main
relation used in \cite{Fucitoiinni} to compute the prepotential.
In the present notation that relation becomes:
\begin{equation}
q\frac{\partial {\cal F}}{\partial q}=-{\bar k}\epsilon_1\epsilon_2
\label{relprep}
\end{equation}
where $\bar k$ is the quantity defined in Eq.~(\ref{domvaluek}).
Using (\ref{prepo}) one finds that
\begin{equation}
\frac{q}{\epsilon_1}\frac{\partial {\cal F}}{\partial q}=
\frac{L^2\log q}{4\epsilon_1}-
\frac{1}{4\epsilon_1}\left[
\sum_{l=1}^{N_c}\left(a_l^2+(a_l-\epsilon_1)^2
\right)-\sum_{r=1}^{N_f}m_r^2
\right]
\end{equation}
The comparison with Eq.~(\ref{normcondtwo}) shows that
the right hand side of the above equation coincides with $-X$ and thus
we have proved the relation (\ref{relprep}) if we remember that
$\bar k=\frac X{\epsilon_2}$.
Let us note that in performing the derivative of $\cal F$
with respect to $q$
it is crucial the fact that
the parameter $L$ does not depend on this variable. Also the
particular dependence of $\cal F$ on $\log q$ turns out to be
essential in order to
guarantee  the consistency between equations (\ref{normcondtwo}) and
(\ref{relprep}).
\section{Conclusions}
The main result of this work is the derivation
of the  exact solution of the
saddle point equation
extremizing the density of eigenvalues $\rho(\phi)$ in the case of
${\cal N}=2$ gauge field theories with gauge group $SU(N_c)$ coupled
to fundamental matter with flavor number $N_f=2N_c$. The choice 
$N_f=2N_c$ made in Eq.~(\ref{assumptionone}) is mandatory when
applying the method of the Fourier transform to the solution of the
saddle-point equation (\ref{eqmots}). Otherwise, the Fourier
transforms of some of the quantities entering in that equation cannot
be uniquely defined. Within the condition (\ref{assumptionone}) the
entire class of conformal ${\cal N}=2$ gauge field theories with
vanishing $\beta-$function is encompassed. A particular case of such
theories is the $\Omega-$deformed version of the ${\cal N}=2$
supersymmetric QCD with 
gauge group $SU(2)$ and four flavors studied by Seiberg and Witten in
\cite{SW2}. This $\Omega-$deformed theory is connected via the AGT
correspondence to
the classical limit of the 4-point conformal block of Liouville field
theory. It is thus useful in order to compute the accessory parameter
for the Fuchsian uniformization of the 4-punctured sphere.
We would also like to stress that our method for
solving the saddle-point equation (\ref{eqmots}) may  be applied
to the so-called ${\cal N}=2^*$ gauge field theories with $SU(N_c)$
gauge group and a matter hypermultiplet in the adjoint representation
\cite{Fucitoiinni}. 

The expressions of
the two contributions $\rho_q(\phi)$ and $\rho_{Q_0}(\phi)$ to
$\rho(\phi)$ are provided
in Eqs.~(\ref{calcrhoq}) and
(\ref{rhokdifzero}). Let us note that this solution depends on two
parameters, namely the half length $L$ of the interval in which the
one-cut solution $\rho_q(\phi)$ is defined and the parameter $X$ given
in Eq.~(\ref{domvaluek}). The latter parameter plays an important role
in the Seiberg--Witten theory, because it is connected to the
prepotential $\cal F$ of 
the supersymmetric gauge theory through the relation (\ref{relprep}).
The quantity $q\frac{\partial {\cal F}}{\partial q}$
appearing in Eq.~(\ref{relprep})
is also related
to the well known Matone's relation \cite{Matone}.
Both parameters $X$ and $L$ may be extracted from the implicit
relations provided by Eqs.~(\ref{normcondtwo}), (\ref{rhoQzeroinzero})
and (\ref{reltwo}).

As a byproduct of the computation of the eigenvalue density $\rho(\phi)$,
the instanton energy has been derived in the
thermodynamic limit. 
In this limit the instanton energy $W$ can be written as a sum
of three terms, see Eq.~(\ref{instantonenergy}), whose expressions
have been calculated in closed form in
Eqs.~(\ref{wqqfinal}), (\ref{wqQzerofinal}) and
(\ref{wQzeroQzerofinal}).
Let us note that the requirement that the instanton free energy is
finite leads to the quantization condition of Eq.~(\ref{condone}).
This provides a physically relevant motivation to conditions of this
kind that have been postulated in previous publications,
like for
instance \cite{Dorey,Dorey2}.
Finally, we have checked the consistency of the prepotential from
instanton counting $\cal F$ obtained  in
Section 3
exploiting the formula of
 the energy $W$ previously derived.
To perform the check, we have used
the differential equation (\ref{relprep}), which has been already
applied in order to compute $\cal F$ in \cite{Fucitoiinni}.
We have verified that the expression of $\cal F$ given in
Eq.~(\ref{prepo}) satisfies the relation (\ref{relprep}) as expected.

Work is in progress in order to compare our exact solution of the
saddle-point equation (\ref{eqmots}) with the previous results, in
which the configurations $\phi(x)$ extremizing the multi-instanton
energy have been computed in terms of a $q-$expansion.
Within our approach, the multi-instanton energy has been considered
from a field theoretical point of view as a functional of the
density $\rho(\phi)$ instead of
$\phi(x)$. The two extremal configurations $\phi$ and $\rho(\phi)$ 
are related to each other by the identity
(\ref{rhophirel}). Thus, the comparison with the previous results
requires to solve
 Eq.~(\ref{rhophirel}) with respect to $\phi(x)$. 
Before doing that, it is necessary to express the parameter $L$
entering in the solution $\rho(\phi)$ by means of the parameters of
the gauge theory. This may be achieved only numerically by solving
Eq.~(\ref{reltwo}) with respect to $L$.
\section{Acknowledgments}
This research has been supported in part by the Polish National
Science Centre under Grant No. N202 326240.

\end{document}